\documentclass[12pt,preprint]{aastex}

\begin{document}


\title{A NEW STELLAR ATMOSPHERE GRID AND COMPARISONS WITH \emph{HST/STIS}
CALSPEC FLUX DISTRIBUTIONS} 

\author{Ralph~C.\ Bohlin\altaffilmark{1},
Szabolcs~M{\'e}sz{\'a}ros\altaffilmark{2,3},   Scott W. Fleming\altaffilmark{1}, Karl
D. Gordon\altaffilmark{1,4}, Anton M. Koekemoer\altaffilmark{1}, and J{\'o}zsef
Kov{\'a}cs\altaffilmark{2}} 

\altaffiltext{1}{Space Telescope Science Institute, 3700 San Martin Drive,
Baltimore, MD 21218, USA} 
\altaffiltext{2}{ELTE Gothard Astrophysical Observatory, H-9700 Szombathely,
Szent Imre Herceg St. 112, Hungary}
\altaffiltext{3}{Premium Postdoctoral Fellow of the Hungarian Academy of Sciences}
\altaffiltext{4}{Sterrenkundig Observatorium, Universiteit Gent,
              Gent, Belgium}

\begin{abstract} The Space Telescope Imaging Spectrograph (STIS) has measured
the spectral energy distributions (SEDs) for several stars of types O, B, A, F,
and G.  These absolute fluxes from the CALSPEC database are fit with a new
spectral grid computed from the ATLAS-APOGEE ATLAS9 model atmosphere database
using a chi-square minimization technique in four parameters. The quality of the
fits are compared for complete LTE grids by Castelli \& Kurucz (CK04) and our
new comprehensive LTE grid (BOSZ). For the cooler stars, the fits with the MARCS
LTE grid are also evaluated, while the hottest stars are also fit with the NLTE
Lanz \& Hubeny OB star grids. Unfortunately, these NLTE models do not transition
smoothly in the infrared to agree with our new BOSZ LTE grid at the NLTE lower
limit of T$_{\rm eff}$=15,000K.

The new BOSZ grid is available via the Space Telescope Institute MAST archive
and has a much finer sampled IR wavelength scale than CK04, which will
facilitate the modeling of stars observed by the James Webb Space Telescope
(JWST). Our result for the angular diameter of Sirius agrees with the
ground-based interferometric value. \end{abstract}  

\keywords{stars: atmospheres --- stars: fundamental parameters
--- techniques: spectroscopic}

\section{INTRODUCTION}

Stellar standards with accurate absolute fluxes (irradiance) are required for
the calibration of the James Webb Space Telescope (JWST) and for the
interpretation of dark energy measures with the supernova Ia technique. The JWST
instrumentation and wavelength coverages include the Mid-Infrared
Instrument (MIRI 4.9--28.8 \micron), the Near-Infrared Camera (NIRCAM
0.6--5\micron), the Near-Infrared Spectrometer (NIRSPEC 0.6--5.3 \micron), and
the Near-Infrared Imager and Slitless Spectrograph 
(NIRISS 0.6--5 \micron).

The basis of the calibrated HST absolute fluxes is the set of three spectral
energy distributions (SEDs) of NLTE model atmospheres for the pure hydrogen
white dwarfs (WDs), GD153 and GD71, and a NLTE metal line-blanketed model of
\citet{Rauch13} for G191B2B \citep[]{Bohlin14,Bohlinetal14,bohlin15}. The STIS
net signal in electrons/s from each of these primary standard WDs relative to
the STIS net signal for Vega at 5557.5~\AA\ (5556~\AA\ in air) defines the
absolute normalization of each model flux. \citet{Bohlin14} reconciled the
5557.5~\AA\ \citet{Megessier95} flux  value with the MSX 8--21~\micron\ fluxes
of \citet{Price04} to establish the absolute flux of $3.44\times10^{-9}$~erg
cm$^{-2}$ s$^{-1}$~\AA$^{-1}$ for Vega at 5557.5~\AA. In the IR, an additional
primary standard is the special Kurucz model for Sirius, which fits both the
5557.5~\AA\ and the average of the MSX measured fluxes to 0.5\%
\citep{Bohlin14}.

The spectral energy distributions of these three primary WDs establish the
calibration of all the HST instruments, including the STIS and NICMOS
spectrophotometers. Model atmospheres can then be fit to HST spectrophotometry
for secondary flux standards; and these model fluxes establish a set of SEDs for
calibrating JWST at IR wavelengths longer than the 1~\micron\ STIS cutoff or the
2.5~\micron\ NICMOS limit. (See Tables~\ref{table:gfits} and \ref{table:ofits}
for the list of our JWST standards and for the results discussed in Section 5.)
Previously, \citet{Bohlin08} fit the \citet[][CK04]{Castelli03} grid to A stars,
and \citet[][B10]{Bohlin10} fit STIS and NICMOS fluxes for G stars with both
CK04 and MARCS \citep{gustafsson08} grids. The stars from these earlier works
are re-fit here because of updated fluxes, improved techniques, and our new
model grid.

The MARCS grid does not cover effective temperatures (T$_{\rm eff}$) above
8000~K; so another grid is required for comparison with the CK04 results. For
our hottest stars, the Lanz NLTE grid \citep[]{lanz03, lanz07} is appropriate,
because the difference between LTE and NLTE becomes significant above
$\sim$15,000~K. Our new grid that is discussed in Section~2 provides full
coverage of our complete sample of stars; and Section~3 details the instructions
for accessing the new grid. Section~4 explains our methodology and highlights
some of our general results, while Section~5 illuminates the pros and cons of
the separate grids. Section~6 summarizes some of the improvements of our new
model SEDs, named BOSZ, using some letters from the names of the first two
authors. Model SEDs from this new BOSZ grid are available via the Mikulski
Archive for Space Telescopes (MAST) at the
STScI\footnote{\url{https://archive.stsci.edu/prepds/bosz}}. For each star, the
best fitting BOSZ model at the full resolution of R=300,000 is extracted from
our BOSZ grid, normalized to the observed SED, and placed in the CALSPEC
database\footnote{http://www.stsci.edu/hst/observatory/crds/calspec.html} along
with a separate file containing the observed SED. From the long wavelength limit
of the STIS or NICMOS data to 32~\micron, this second CALSPEC file is a
concatenation of the observed HST SED with the best fit R=500 model from our new
BOSZ grid and is most appropriate for calibration of low resolution (R
$\lesssim$ 500) data. Table~\ref{table:cfmod} compares the four sets of model
atmosphere grids that are used in this paper.

\begin{deluxetable}{l|l|l|l|l|}		
\tabletypesize{\scriptsize}
\tablewidth{0pt}
\tablecaption{Parameters of the Full Model Grids}
\tablehead{
\colhead{parameter} &\colhead{BOSZ} &\colhead{CK04\tablenotemark{a}} &\colhead{MARCS} 
	&\colhead{Lanz\tablenotemark{d}}}
\startdata
Type &	LTE   & LTE  & LTE & NLTE  \\
Wavelength range (\micron) & 0.1~--~32 & 0.09~--~160&0.13~--~20&0.0045~--~300\\
T$_{\rm eff} (K)$  & 3500~--~30,000 & 3500~--~50,000&2500~--~8000&15,000~--~55,000 \\
$\log~g$  &0~--~5 See Table 2 & 0~--~5 See CK04&-1~--~5\tablenotemark{b}&1.75~--~4.75\tablenotemark{e}\\
$[M/H]=\log z$	   &-2.5 to 0.5&-2.5 to 0.5&-5 to 1\tablenotemark{c}&-1 to 0.3  \\
$[C/M]$	   & -0.75 to 0.5 & 0 & 0\tablenotemark{f} & 0 \\
$[\alpha/M]$	   & -0.25 to 0.5 & 0, +0.4 & 0\tablenotemark{f}  & 0  \\
Solar Abundance&\citet{asplund05}&\citet{grevesse98}&\citet{grevesse07}&\citet{grevesse98}   \\
Microturbulent velocity (km/s)&2 & 2 &  0,1,2,5  & 2 ($\leq$30,000K),10($\geq$32,500K) \\
Rotational broadening (km/s) &0 & 0 & 0  & 0 \\
Spectral resolution R&R=200--300,000\tablenotemark{g}& 1221 bins & R=20,000 & R$\sim$1800  \\
Convective mixing length& 1.25 & 1.25 & 1.5  & No convection \\
Convective overshoot& No & No & No & No convection  \\
Continuum          & Yes & Yes & No  & No \\
\enddata
\tablenotetext{a}{As updated from http://wwwuser.oats.inaf.it/castelli/.}
\tablenotetext{b}{Models are plane-parallel, except sphericial
	below $\log~g=3$. There are many missing models due to convergence
	problems, especially for the
	spherical cases. For example, 52 missing models are filled by
	interpolation in the plane-parallel set.}
\tablenotetext{c}{Marcs models have a somewhat different
	definition of $[M/H]$, so that metallicity results for MARCS are not
	directly comparable to the other model results.}
\tablenotetext{d}{As provided in the merged Cloudy format file
	obstar\_merged\_3d.ASCII from 
	http://nova.astro.umd.edu/Tlusty2002/tlusty-frames-cloudy.html.}
\tablenotetext{e}{See \citet{lanz03} and \citet{lanz07} for details.}
\tablenotetext{f}{See \citet{gustafsson08} and http://marcs.astro.uu.se for 
details.}
\tablenotetext{g}{The BOSZ model grid offers 10 choices in the R=200--300,000
range.}
\label{table:cfmod} \end{deluxetable}

\section{THE NEW GRID}  

A publicly available and consistently calculated database of model SEDs is
important for many astrophysical analyses, including spectroscopic surveys and
analysis of elemental composition of stellar atmospheres. Different spectral
grids are available in the literature, and one of the largest is based on ATLAS9
\citet{kurucz79} calculations \citep{zwitter04} for the GAIA mission. More
recently, \citet{coelho14} published a synthetic spectral grid containing 12
compositions covering the metallicity range $[M/H]=\log z$ of -2.77 to -1.31,
where z is the abundance ratio to hydrogen by number of the total sum of all
elements except helium and hydrogen. These metal abundances are relative to the
solar abundances of \citet{asplund05}, where a model with $[M/H]=0$ has the 
solar abundance.
Other libraries of model atmosphere grids are also available in the literature;
for a complete list, see the introduction by \citet{coelho14}. Older ATLAS9
libraries were synthesized using the solar reference abundance table from either
\citet{grevesse98}, or \citet{anders1989} and are limited to a handful of
mixtures. However, in the early 2000s, significant changes and improvements were
made to the solar compositions table \citep{asplund05, grevesse07, asplund09}
that should be used to calculate grids of updated stellar spectra. Two extensive
modern grids, CK04 and MARCS, use the \citet{grevesse98} and \citep{grevesse07}
abundance tables, respectively.

\begin{deluxetable}{lrrrlrrr}	
\tabletypesize{\scriptsize}
\tablecaption{Atmospheric Parameters of ATLAS9 Spectra}
\tablewidth{0pt}
\tablehead{
Parameter & \colhead{Min} & \colhead{Max}    & \colhead{Step}  &  
Parameter & \colhead{Min} & \colhead{Max}    & \colhead{Step}
}
\startdata
$[$M/H$]$ & $-$2.5 & 0.5 & 0.25 & & & &\\
$[$C/M$]$ & $-$0.75 & 0.5 & 0.25 & & & & \\
$[\alpha$/M$]$ & $-$0.25 & 0.5 & 0.25 & & & & \\
T$_{\rm eff}$  & 3500 & 6000 & 250 & $\log$~g & 0 & 5 & 0.5 \\
T$_{\rm eff}$  & 6250 & 8000 & 250 & $\log$~g & 1 & 5 & 0.5 \\
T$_{\rm eff}$  & 8250 & 12000 & 250 & $\log$~g & 2 & 5 & 0.5 \\
T$_{\rm eff}$  & 12500 & 20000 & 500 & $\log$~g & 3 & 5 & 0.5 \\
T$_{\rm eff}$  & 21000 & 30000 & 1000 & $\log$~g & 4 & 5 & 0.5 \\
\enddata
\label{table:params} \end{deluxetable}

\begin{figure}[!ht]	
\epsscale{0.75}
\plotone{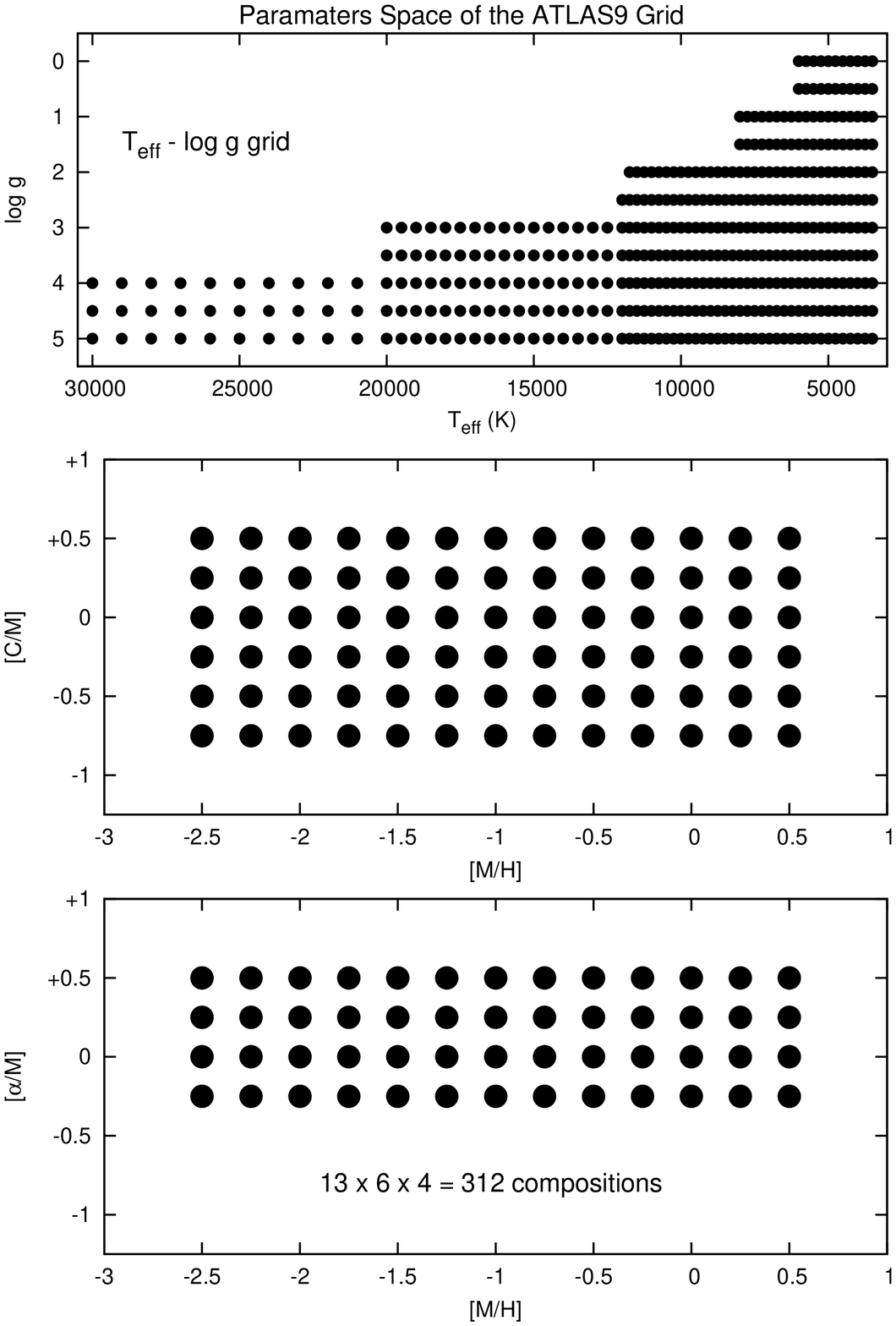}
\caption{Top panel: The T$_{\rm eff} - \log$~g space of model atmospheres used
for synthesis.  Middle panel: [C/M] abundances as a function of
metallicity. Bottom panel: [$\alpha$/M] abundances for each [C/M]
values as a function of metallicity. The product of our 13 [M/H], 6 [C/M], and 
4 [$\alpha$/M] values comprises the full parameter space of the 312 compositions.
\label{synspace}} \end{figure}

Recently, \citet{meszaros12} calculated ATLAS9 LTE model atmospheres for a
larger number of compositions. Using a 2~km/s microturbulent velocity, this grid
covers metallicities from [M/H]=$-5.0$ to [M/H]=+1.0 with carbon ([C/M]) and
$\alpha$ ([$\alpha$/M]) abundances from $-$1.5 to +1.0 relative to solar
metallicity. The $\alpha$ elements (i.e. those with an even number of protons)
that are varied in these calculations are O, Ne, Mg, Si, S, Ca, and Ti. One of
the most important differences between \citet{meszaros12} and previous model
atmosphere grids is the use of more recent Solar abundances from
\citet{asplund05}; but the \citet{meszaros12} output grid SEDs contain only
333 wavelength points.

This plane-parallel dataset of \citet{meszaros12} is the model atmosphere grid
for spectral synthesis of a new library that spans most of the atmospheric
parameter space that is known to exist for actual stars. Atlas9 is chosen,
instead of the available Atlas12 Kurucz code, because of the large variety of
compositions in the grid and because Atlas12 requires more computation time and
often fails to converge. The high-resolution spectra are calculated from 
the \citet{meszaros12} model atmosphere grid with SYNTHE
\citep{kurucz81} using the Linux-ported version \citep{sbordone04}. The selected
atmospheric parameters are listed in Table~\ref{table:params} and shown in
Figure~\ref{synspace}. With a total of 312 different compositions that are
selected to cover the majority of actual stellar abundances, our 13 [M/H], 6
[C/M], and 4 [$\alpha$/M] values are a subset of the original \citet{meszaros12}
grid. The spectra span the wavelength range of 1000~\AA~--~32~\micron\ using
vacuum wavelengths and are first synthesized with one sample point per
resolution element at R~=~300,000 without convective overshooting and with a
mixing length parameter of 1.25. Lower-resolution spectra are produced by
convolving the highest resolution SED with Gaussian line-spread functions of
full-width-at-half-maximum $\lambda/R$. Nine additional resolutions are provided
with two points per resolution element: R = 100,000, 50,000, 20,000, 10,000,
5,000, 2,000, 1,000, 500, and 200. The computed fluxes are sampled evenly in
logarithmic wavelength space with a sample spacing of $\lambda/2R$ for two
points per resolution element, and all SEDs with the same resolution have the
same common set of wavelengths. No rotational broadening is applied to any of
the models, while the 2 km/s micro-turbulent velocity is the same as used in the
model atmosphere computation. Details of the naming convention for the MAST
models are in the Appendix.

The atomic line list compiled by Robert
Kurucz\footnote{http://kurucz.harvard.edu/linelists.html} is used without
modification and is complemented with the molecular line lists for H$_{2}$O, CH,
MgH, NH, OH, SiH, H$_{2}$, C$_{2}$, CN, CO, SiO, and
TiO\footnote{http://kurucz.harvard.edu/molecules.html}. These line lists are 
frequently updated, but the April 2015 version is used for all the calculations.
The H$_{2}$O \citet{partridge97} and TiO \citet{schwenke98} line lists were
formatted for ATLAS9 by Robert Kurucz. Water is included for stars cooler than
5500~K and TiO only below 4500~K to reduce computation time for temperatures
where these molecules are unimportant. The computations are parallelized by
composition. Each composition has 61 different temperatures and 3--11 different
gravities according to Table~\ref{table:params}, i.e. 415 models per composition
and 312*415=129480 total models. Each model SED is available at the 10 different
spectral resolutions for a total of 4150 files per composition and a grand total
of 1.3 million files in the complete BOSZ grid.

\section{THE GRID ARCHIVE} 

Each model SED has three columns: wavelength in \AA, surface brightness
\textit{h(BOSZ)}, and theoretical continuum level in the same units as
\textit{h(BOSZ)}.  While theoretical models often are in units of Eddington flux
\textit{H} where the flux at the stellar surface is 
\begin{equation}{F=4~\pi~H,}\end{equation} e.g. Sirius from the Kurucz web 
site\footnote{http://kurucz.harvard.edu/stars/sirius/sirallpr16.500resam501} and
\citet{tremblay2017}. Our \textit{h(BOSZ)} values are four times larger than H,
so that \begin{equation}{F=\pi~h(BOSZ).}\end{equation} The units of our BOSZ
flux F emitted at the stellar surface is erg cm$^{-2}$ s$^{-1}$ \AA$^{-1}$,
while the Kurucz flux unit is ten times smaller, i.e. erg cm$^{-2}$ s$^{-1}$
nm$^{-1}$. In order to properly account for limb darkening, the original model
calculations produce specific intensity \textit{I} at 17 angles with respect to
the stellar surface. Both \textit{h(BOSZ)} and \textit{H} are numerical
integrals of \textit{I} over 2$\pi$ steradians and represent the net out-going
flux at optical depth zero according to Equations 1 and 2.

In the absence of interstellar extinction, the total stellar luminosity
in erg s$^{-1}$ \AA$^{-1}$
is the same at the stellar radius \textit{R} and at the distance to the Earth
\textit{r}

\begin{equation}{4~\pi~R^2~F=4~\pi~r^2~f,}\end{equation}
where \textit{f} is the
measured absolute flux distribution. Thus, the stellar angular diameter is 
\begin{equation}{\theta=2R/r=2~\sqrt{f/F}.}\end{equation}

\begin{figure}			 
\centering  
\includegraphics*[width=1\textwidth,trim=0 10 0 0]{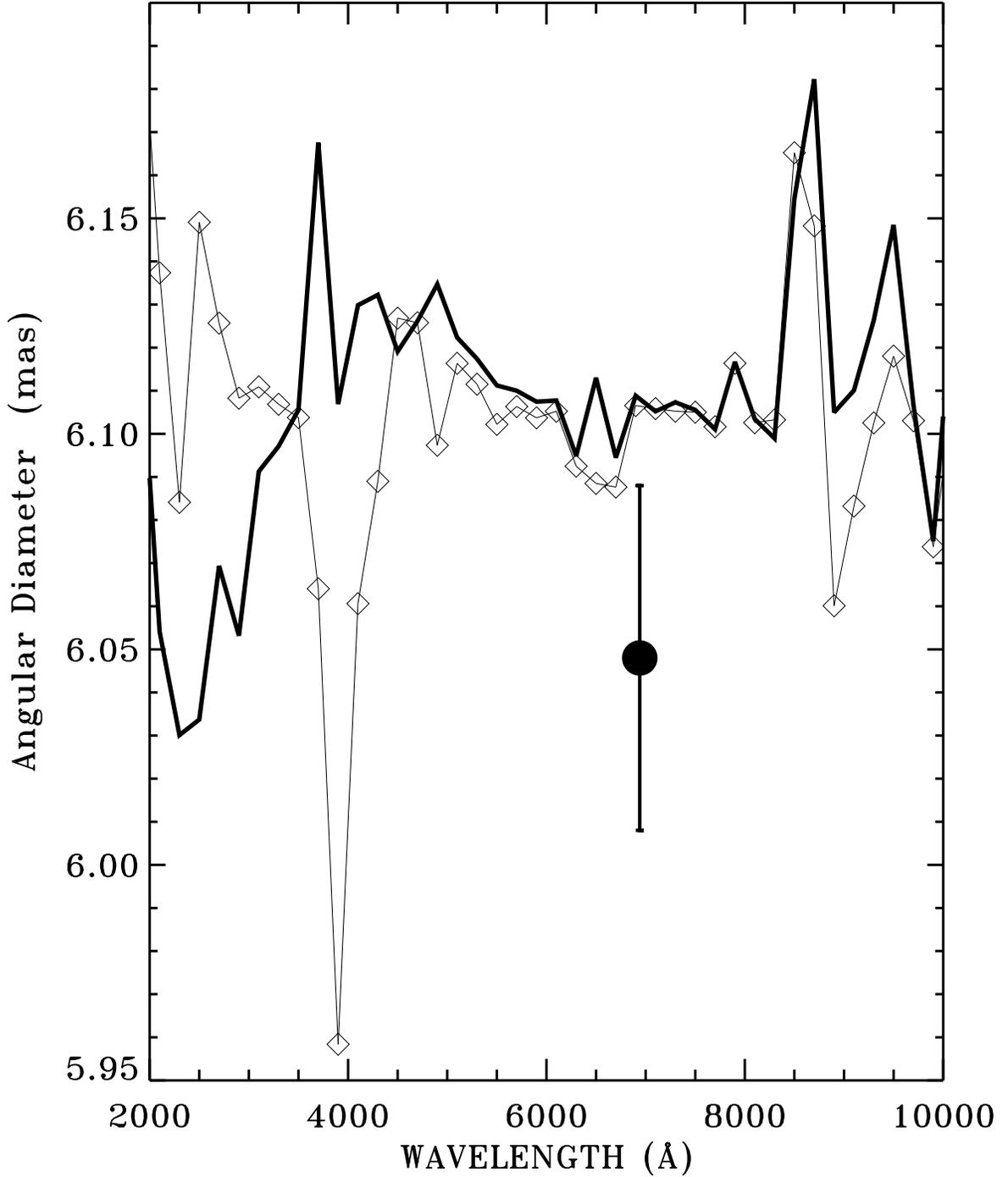}
\caption{\baselineskip=12pt
Angular diameter of Sirius as a function of wavelength in 200~\AA\ bins.
The heavy solid line
utilizes the specially constructed model from the Kurucz website; and the thin
line with diamonds represents the best fit of our BOSZ models with 
$T_\mathrm{eff}$=9820~K, $\log g$=3.95, $[M/H]$=0.45, and E(B-V)=0.000. The
interferometric measure of \citet{davis2011} at 6941~\AA\ is the filled circle
with error bar.
\label{angdiam}} \end{figure}

Figure~\ref{angdiam} is an example of the use of the measured CALSPEC flux
\textit{f} for Sirius \citep{Bohlin14} along with model atmosphere calculations
\textit{F} to determine the angular diameter. The heavy solid line represents
the angular diameter as a function of wavelength using the Kurucz model for
\textit{H} with  $T_\mathrm{eff}$=9850~K, $\log g$=4.3, and $[M/H]$=0.4, while
the light line with diamonds is the result of fitting the CALSPEC SED with our
\textit{h(BOSZ)} grid (see the next section.) The BOSZ fit provides a slightly
more constant value of $\theta$, except in the region of the Balmer line
confluence, where the exact opacity is not known. Different modeling codes use
different approximations to calculate this opacity. The data point is the
interferometric measurement of 6.048~mas \citep{davis2011} with its 0.04~mas
error bar. (Also compare with the discussion in \citet{linnell2013}.) While the
\citet{davis2011} measure agrees with the two results from both models within
the 1$\sigma$ uncertainty, increasing the model surface flux or decreasing the
measured absolute CALSPEC flux by $\sim$1\% would reduce the difference.
\citet{davis2011} also discuss other measurements of the angular diameter of
6.039 and 6.089 at 2.2~\micron\ both with a 0.02 mas uncertainty but with
questions raised about the validity of the uncertainty for the 6.089 value.

The total volume of the bz2 zipped ASCII version of our new grid is
$\sim$3.7~TB, and access to grid subsets is at the MAST High Level Science
Product (HLSP) page for the BOSZ models. 

Models can be retrieved in a variety of ways.  A web search form allows users to
specify one or more parameter selections among $T_{\rm{eff}}$, $\log(g)$, [M/H],
[C/M], [$\alpha$/M], and/or instrumental broadening.  For relatively small
numbers of models, users can directly download the ASCII or FITS versions from
links in the search results table.  For larger numbers of models, users can opt
to obtain a wget or cURL script as output, which they can save and then execute
on their own machines for sequentially downloading each of the requested files. 
Users can also download pre-constructed bundles of models for commonly requested
sets (e.g., all models at a particular instrumental broadening, or all models at
solar relative abundances but differing overall metallicity).  For detailed
instructions, consult the MAST HLSP page for BOSZ.

\section{FITTING MODELS TO OBSERVED SEDs} 

The model that best fits the observed CALSPEC SED is determined by the minimum
chi-square fit of models interpolated from a model atmosphere grid. A set of
wavelength bins are selected and the chi-square ($\chi^2$) difference between
the data and a model for each bin is computed along with the reduced chi-square
average, where the reduction is by the four free parameters of the fit. The
expected uncertainties for the $\chi^2$ calculations depend on the amount of
line-blanketing in the models, the background noise of the observations, and the
broadband repeatability of the observed CALSPEC SEDs for bright stars where the
sky background is negligible. The global search for a minimum chi-square
proceeds over the four parameters $T_\mathrm{eff}$ (10), $\log g$ (0.05), $\log
z$ (0.01), i.e. [M/H], and interstellar reddening from the dust E(B-V) (0.001),
where the step size of the search in each parameter is in parentheses. With a
total to selective extinction ratio at V of 3.1, the extinction curve for
interstellar reddening is from \citet{cardelli89} below 2~\micron\ and from
\citet{chiar06} above 2~\micron, where the \citet{chiar06} curve is normalized
to \citet{cardelli89}. This composite extinction is multiplied by the selective
extinction E(B-V) to get the total extinction $A_\lambda$ in magnitude units.

The fitted models provide an estimate of IR fluxes beyond the long-wavelength
limit of the observation; and the established IR flux standards provide
calibration sources for JWST and other IR instrumentation. To limit systematic
effects in the modeling and extrapolation process, a variety of stellar types
are used for an instrumental flux calibration; and averaging the instrumental
sensitivities over several stellar standards of each type provides a statistical
reduction of the random errors in the measured fluxes and in the fitting
process. For example, observing four stars with a 1\% statistical uncertainty
each will reduce the uncertainty to 0.5\% and will provide a flux calibration on
the HST flux scale within the 0.5\% statistical uncertainty. Stars cooler than G
type are avoided because of the added complication of accounting for the
plethora of molecular opacities in their atmospheres.

Table~\ref{table:gbins} lists the wavelength intervals used for fitting G--F
stars, while Table~\ref{table:obins} contains the A--O star bins. All bins have
equal weight in the fitting process, but the weight of diferent wavelength
regions is controlled by the number of bins in each region. The ranges are
chosen to avoid some of the stronger absorption lines that may be less
accurately modeled. Some UV regions with higher line blanketing are avoided, and
the UV regions are generally de-emphasized because of the severe blanketing that
makes the models less precise. However to some extent, averaging over broad bins
reduces the statistical uncertainties of absorption line strengths in the
models. Bins not covered by observed CALSPEC STIS or NICMOS fluxes are omitted,
and UV bins with average flux $<$1\% of the peak stellar flux are also not
considered. The 1--1.3~\micron\ region is avoided because of larger
uncertainties in the NICMOS non-linearity correction of \citet{bohlin06}. The
two bins at the longest wavelengths are not used for P177D, because there is
only a single NICMOS G206 observation.

\begin{figure}[tbp] 
\epsscale{0.85}
\plotone{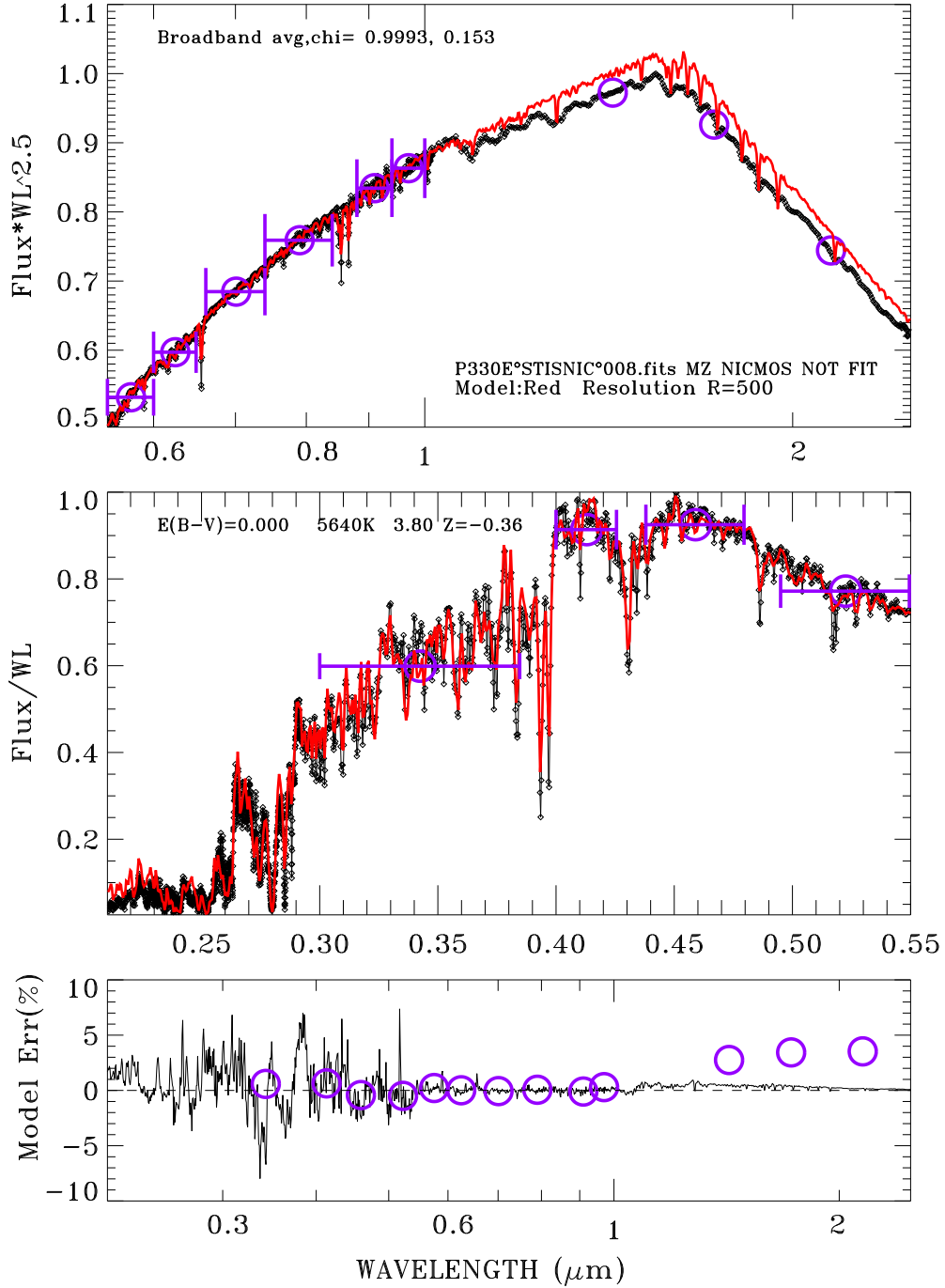}
\caption{
BOSZ fit to only the CALSPEC STIS data for P330E. In the top two panels, black
is the measured SED with STIS data below the 10100~\AA\ crossover point with
NICMOS, while the red model has the parameters on the middle panel. Both model
and data are scaled by the same power of wavelength according to the Y-titles.
Purple circles represent the data averages in the bins of
Table~\ref{table:gbins} and horizontal error bars are the bin widths.
Circles without error bars are the NICMOS averages that are not used in the
fitting of the model. The bottom panel shows the residuals to the fits where the
black line is the difference between model fit and data on a scale of $\pm10\%$
of the peak stellar flux. The purple circles are the average binned differences
on a scale of $\pm10\%$ of the local flux. \label{findmin}} \end{figure}

\begin{figure}	
\centering 
\includegraphics*[width=1\textwidth,trim=0 0.1 0 0]{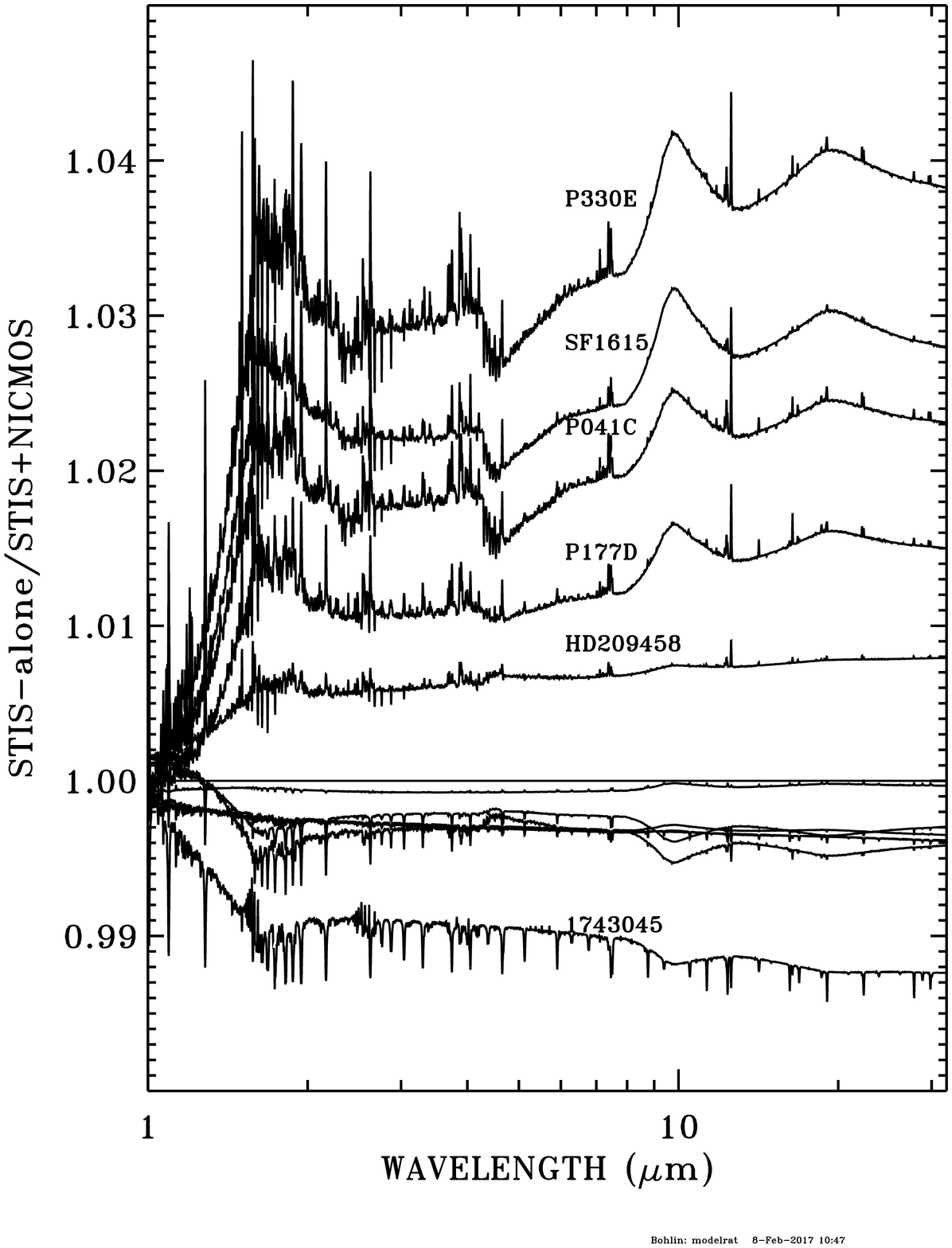}
\caption{\baselineskip=14pt
Ratios of two independently derived IR SEDs of BOSZ fits to our 13 stars 
with NICMOS data. The
numerators consider the fit to only the STIS portion below 1~\micron, and the
denominators are the model fits to the full STIS+NICMOS datasets that extend to
2.5~\micron. The ratios for the A stars 1802271 and 1812095 fall at unity;
and the six outlier ratios are labeled. The remaining two G stars (C26202 and
SNAP2) and three A stars (1732526, 1805292, and HD165459)
fall between 0.995 and 1.000.
\label{modelrat}}
\end{figure}
	
While all of our BOSZ models are available from the MAST Archive, only models
with scaled relative solar abundances are considered here, in order to have a
proper comparison with the other solar abundance grids discussed in this paper.
Possible variations of carbon and $\alpha$-process elements are deferred to a
separate study of fitting with the full BOSZ model composition set. Our
technique finds the best $\chi^2$ fits for the wavelength bins of Tables 2--3.
Because many stars do not have the NICMOS IR coverage beyond the STIS 1~\micron\
limit, a comparison of the model fits is made with and without the NICMOS data
in order to assess the possible errors for the stars that have only STIS data.
The R=500 BOSZ models that approximately match the STIS resolution are used. For
the seven G stars and six A stars with NICMOS coverage, the models from the
STIS-only fits all agree with the measured NICMOS SEDs within the $\sim$2\%
NICMOS uncertainty in all three IR bins, except for P330E and SF1615+001A, where
the errors exceed 2\% at the 2.5~\micron\ long wavelength limit for NICMOS. For
the worst case of P330E with our new model grid, Figure~\ref{findmin}
illustrates the STIS-only fit and the residuals.  Figure~\ref{modelrat}
illustrates the IR ratios of the best fit BOSZ models without and with
consideration of NICMOS. The ratio for P330E in  Figure~\ref{modelrat}
represents a worst case (perhaps 3$\sigma$) uncertainty of our extrapolated IR
SEDs. The results for the CK04 and MARCS grids are similar. The BOSZ fit
parameters to the P330E STIS+NICMOS data, i.e.  $T_\mathrm{eff}$=5840,  $\log
g$=4.40, $[M/H]$=-0.16, and \mbox{E(B-V)=0.036}, are from
Table~\ref{table:gfits}, while the STIS-only fit parameters are
$T_\mathrm{eff}$=5640, $\log g$=3.8, $[M/H]$=-0.36, and E(B-V)=0.000, i.e. 200~K
cooler. This case reflects the partial degeneracy between  $T_\mathrm{eff}$ and
reddening E(B-V). Without the constraints of  a robust measure of a UV continuum
from 1500--3000~\AA, a higher $T_\mathrm{eff}$ is compensated by a higher
reddening that produces the same model flux to $<$1\% in all of our G-star bins
in the STIS wavelength range. Even with the constraint of the NICMOS SED to
2.5~\micron, the chi-square values for the 13 bins of Table~\ref{table:gbins}
range from only 0.1 to 1, which reflects the underconstrained nature of the
fitting process. The UV fluxes for G stars are not used as constraints, because
even the solar flux is known to vary by more than a percent below 2500~\AA.
Longer wavelength IR photometry, such as Spitzer/IRAC data would provide
additional very helpful contraints. 

Nevertheless, the case of P330E most likely represents an extreme error and not
a systematic problem with our fits to the STIS-only SEDs. In general, the NICMOS
data confirm the STIS-only model fits to $\sim$2\% at the 2.5~\micron\ long
wavelength limit for most of the stars with NICMOS spectra. Thus, the expected
1$\sigma$ uncertainty of our modeled IR fluxes should be of order 1\% with
respect to the HST flux scale. However, the G stars in Figure~\ref{modelrat}
include the three outliers P330E, SF1615+001A, and P041C, which are all
$\sim$2\% or more high. Thus as a group, the A stars are more reliable
IR flux standards than the G stars; and IR standards with NICMOS data are 
preferred over those with only STIS data, especially in the case of G stars.

\section{THE MODEL FITS} 

\subsection{G Stars}

\begin{figure}[tbp]	
\epsscale{0.85}
\plotone{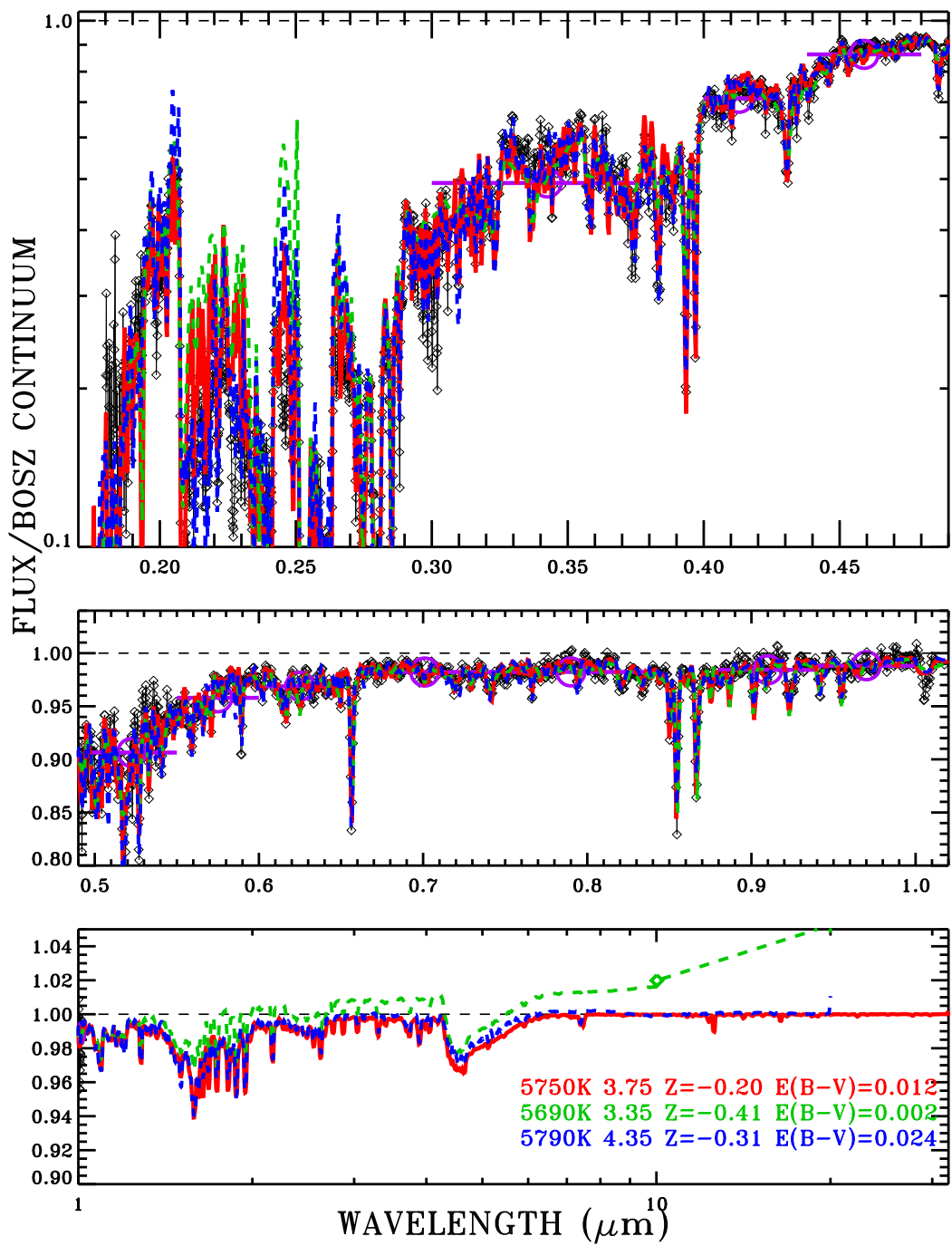}
\caption{
The best fitting models for HD37962, where our BOSZ model is red, green dashes
represent the CK04 model, and the blue dashed line is the MARCS model. The STIS
data are the small black diamonds connected with a thin black line.
In the bottom panel, the green
diamond point at 10~\micron\ is one of only three CK04 sample points in the
entire 10--40~\micron\ range.
\label{modlcont}} \end{figure}

Table~\ref{table:gfits} includes the results for 12 G stars using bins longward
of 3000~\AA. Over the fitted range, all three grids produce comparable results
often with the same $T_\mathrm{eff}$ to 40~K. From 1-20~\micron, the BOSZ and
MARCS model fits always agree to $\sim$1\%, while the CK04 fits
occasionally disagree with the BOSZ by nearly 2\%. A worst case is HD37962,
where the CK04 fit approaches 2\% higher than the BOSZ fit at 10~\micron, as
shown in Figure~\ref{modlcont}. From 10--40~\micron\, there are only three
tabulated CK04 flux values, which are not very useful for comparing with
observation in the mid-IR. 

\begin{figure}			 
\centering  
\includegraphics*[width=1\textwidth,trim=0 0.1 0 0]{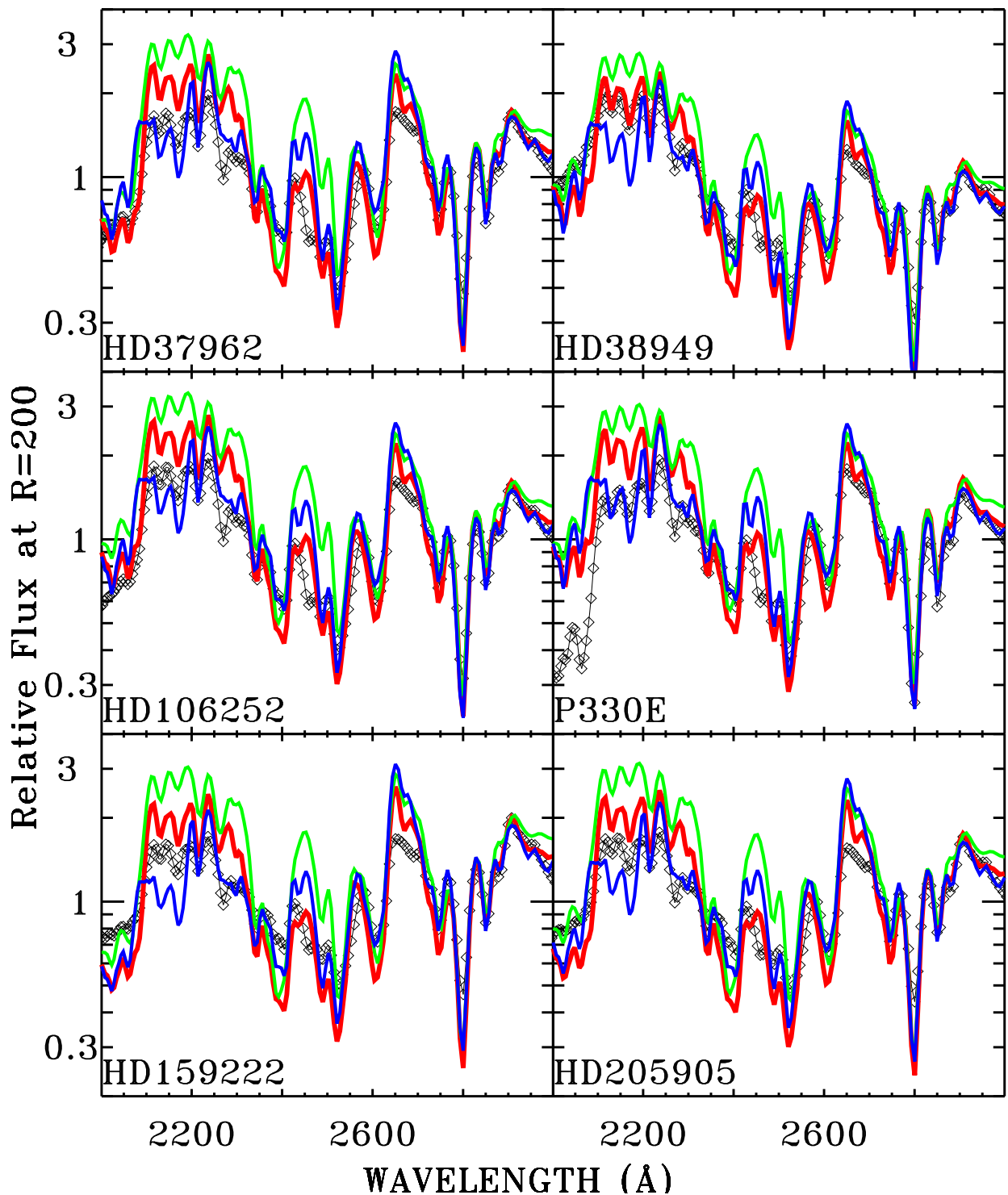}
\caption{\baselineskip=12pt
Black diamonds: STIS UV fluxes for six G stars. Models: Red--BOSZ,
Green--CK04, Blue--MARCS. All three models from
Table~\ref{table:gfits} and the STIS data are smoothed to a resolution R=200.
The models are
normalized to STIS at 6800--7700\AA. Both data and models are
divided by the wavelength to the 8th power for display. 
\label{uvsed}} \end{figure}

The six G stars with UV data below 3000~\AA\ are compared to the short
wavelength extensions of their models in Figure~\ref{uvsed}. Below 3000~\AA, 
line-blanketing removes over half of the continuum flux, making modeling
problematic because of our incomplete knowledge of the physical parameters of
the atomic lines. Thus, occasional deviations of the models from the data by a
factor of two is not surprising. However, the quality of the fits might depend
on deviations of the carbon and $\alpha$-process elements from the solar
abundance assumed for the models in Figure~\ref{uvsed}.

No model matches the data to 20\% over the entire 2000--3000~\AA\ range of 
Figure~\ref{uvsed}. However, the MARCS models match the data from
2700--3000~\AA\ and to 30\% over most of the range, except for
P330E. The quality of our BOSZ fits is comparable to MARCS, while
the CK04 fits tend to be the outliers.
All three models are
always higher than the data in the 2650~\AA\ region for all six stars, although
the BOSZ models have only about half the error of the other two fits.

\subsection{A--O Stars}		

Table~\ref{table:ofits} includes the results for both our BOSZ and the CK04
grids for the 19 stars that are hotter than 7000~K. The $T_\mathrm{eff}$ results
for the two cases all agree to 40~K, except for HD158485 with a difference of
60~K. There are six stars with $T_\mathrm{eff}$ below the MARCS upper limit of
8000~K; and the MARCS fits are systematically cooler than our BOSZ results by
by 50--100~K. This systematic difference just means that the same SED shape is
labeled as cooler in this temperature range of the MARCS grid, but both grids
provide equally valid results.
The models in the line-free regions agree to 1\% for the
comparison of BOSZ to CK04 out to 10~\micron, while the BOSZ and MARCS fits also
agree to 1\% to the 20~\micron\ limit of the MARCS grid. This consistency
reinforces the above suggestion that our BOSZ extrapolations have a 1$\sigma$
uncertainty of order 1\% in the IR.

\begin{figure}[tbp] 
\epsscale{.9}
\plotone{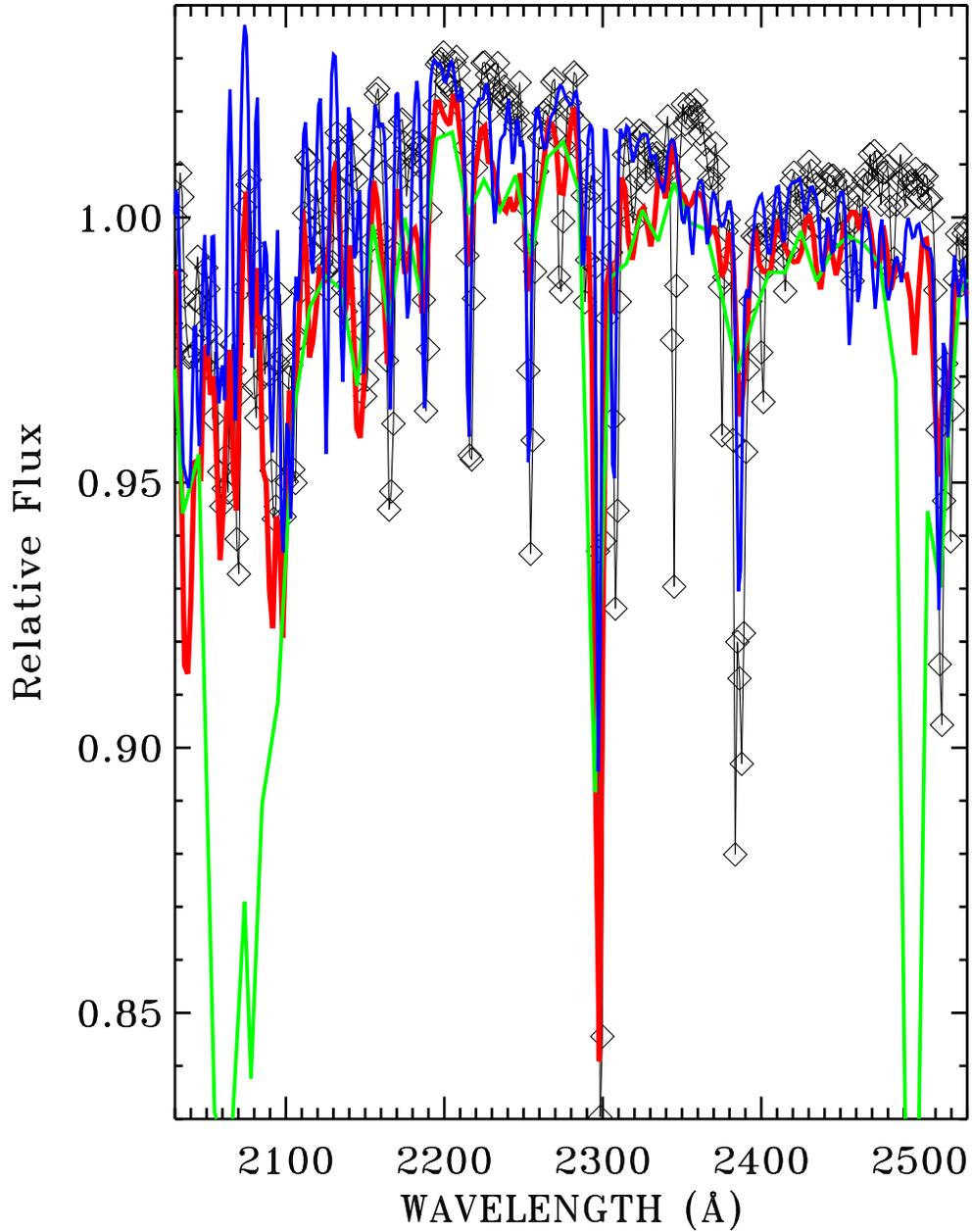}
\caption{
STIS and modeled fluxes for $\mu$ Col as in Figure~\ref{uvsed}, except that the
resolution is R=500 and the scaling is by wavelength cubed. The STIS data are
the black diamonds connected with a thin black line. Our BOSZ model is
red, and the blue line is from the \citep{lanz07} B-star
grid. The solid green line illustrates the
two erroneous CK04 features.\label{pltsed-bad}} \end{figure}

At higher $T_\mathrm{eff}$, \citet{lanz03} for $T_\mathrm{eff}$ 27,500--55,000~K
and \citet{lanz07} for $T_\mathrm{eff}$ 15,000--30,000~K compute NLTE model
grids, which better represent the physics of the hottest stellar atmospheres
than LTE models. These NLTE Lanz models are binned in frequency and merged into
the CLOUDY format in the file
$obstar\_merged\_3d.ASCII.gz$\footnote{nova.astro.umd.edu/Tlusty2002/tlusty-frames-cloudy.html}.
Above $T_\mathrm{eff}=15,000$, the NLTE models characterize our three O stars,
where the best fits are summarized in Table~\ref{table:lanz}. The modeled NLTE
strengths of Paschen absorption lines are particularly sensitive to the
parameters of the fits, so two wavelength bins at 9549 and 9232~\AA\ (vacuum)
for P$\epsilon$ and P9 are included in Table~\ref{table:obins} in order to
improve the fits to the Paschen equivalent widths. Not only does the line
structure often match the data better in the Figure~\ref{pltsed-bad} example,
but also the $\chi^2$ is up to a factor of two better for the Lanz OB-star
grid. 

\begin{figure}[tbp] 
\epsscale{0.85}
\plotone{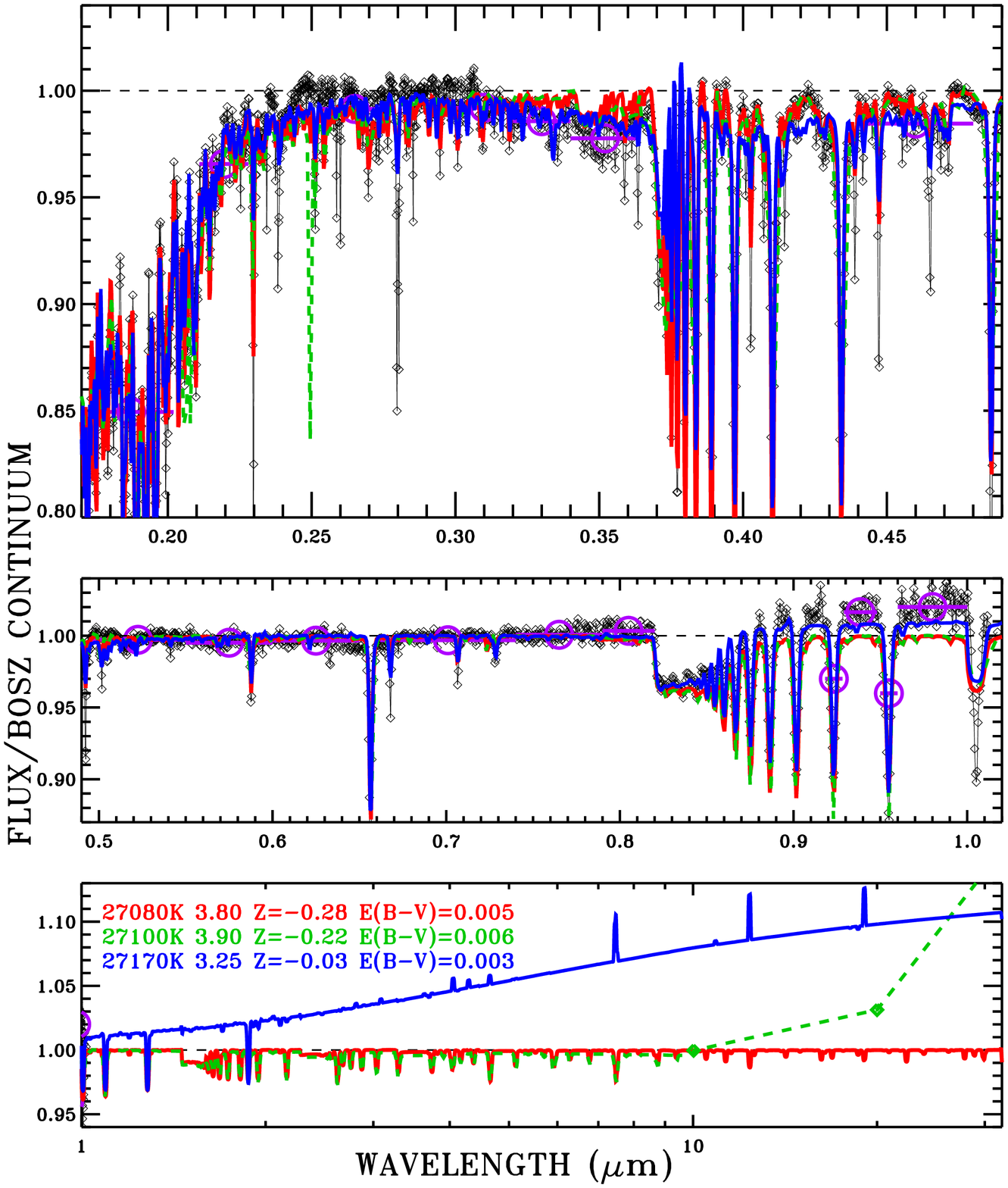}
\caption{
As in Figure~\ref{modlcont} for $\lambda$ Lep, where 
the STIS data are
the black diamonds connected with a thin black line, our BOSZ model is
red, CK04 is green, and the blue line is for the Lanz B-star NLTE grid.
Despite agreement of the data with all three
models to $\sim$2\% in broad bins, the NLTE grid fit differs systematically 
from the two LTE grid fits in the IR. The NLTE model fits the STIS
SED significantly better in the continuum between the Paschen lines at
0.9--1\micron.
\label{modlep}} \end{figure}

\begin{figure}[tbp] 
\epsscale{1.}
\plotone{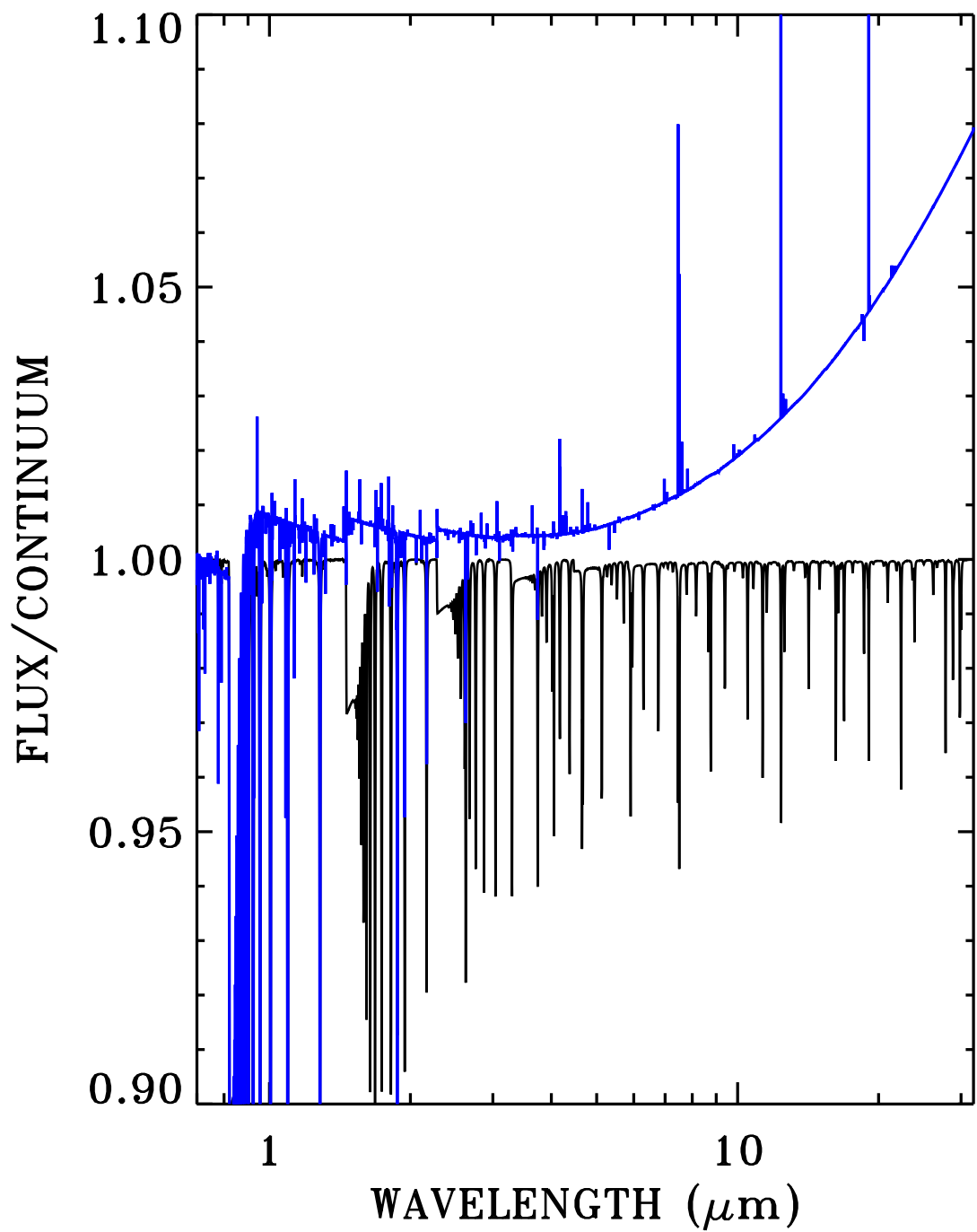}
\caption{
Comparison in the IR of a Lanz NLTE model (blue) at
15,000~K to the best fitting a BOSZ LTE model (black). 
The two models are normalized at 6900--7700~\AA\ and are matched in
the bins of Table~\ref{table:obins}. 
Both models are divided by the BOSZ continuum, which is normalized by the
same factor as the model normalization. 
Unfortunately, the NLTE model is systematically more than 1\% brighter
longward of 5~\micron\ and rises to almost an 8\% deviation at 32~\micron.
\label{ltenlte}} \end{figure}

For the three stars with $T_\mathrm{eff}$ above the 15,000~K lower limit of the
OB-star NLTE grid, the NLTE IR fluxes longward of 1~\micron\ are systematically 
higher than the LTE grid fits. The worst case IR difference is for $\lambda$ Lep
as illustrated in Figure~\ref{modlep}, where the best fitting BOSZ and Lanz
models differ by 11\% at 32~\micron. While there are no CALSPEC stars near the
15,000~K lower limit of the Lanz grids, a proper comparison beween LTE and NLTE
at this cross-over point is accomplished by fitting the Lanz
$T_\mathrm{eff}$=15,000, $\log g$=3.5, and $[M/H]$=0 model with our BOSZ grid
over our usual range of Table~\ref{table:obins}. The best BOSZ fit is 
$T_\mathrm{eff}$=14,980, $\log g$=4.0, and $[M/H]$=-0.36. While the two models
differ slightly in their parameter labels, the important difference is in the IR
fluxes, as illustrated in Figure~\ref{ltenlte}. A comparison using the same
$T_\mathrm{eff}$=15,000, $\log g$=3.5, and $[M/H]$=0 BOSZ model shows a slightly
worse discrepancy. Unfortunately, the NLTE model has systematically higher flux
in the IR with an 8\% divergence at 32~\micron. Either there is some deficiency
in the LTE or NLTE model calculations, or NLTE is important in the mid-IR below
15,000~K. One suspicious aspect of Figure~\ref{ltenlte} is that the NLTE model
shows no Brackett, Pfund, or higher hydrogen line series converging to their
continuum discontinuities like both models at the 0.82~\micron\ Paschen limit.
Perhaps, the problem is just that the NLTE Lanz models utilize an incomplete
model hydrogen atom. In any case, there is evidence that LTE models have no
problem below 10,000~K, where \citet{Bohlin14} found one percent average
agreement between a special LTE Kurucz model at 9850~K for
Sirius\footnote{http://kurucz.harvard.edu/stars/sirius/} and MSX absolute flux
measures at 8, 12, 15, and 21~\micron. Futhermore, a fit of BOSZ models to this
special Kurucz model for Sirius is consistent with $<$1\% difference at
32~\micron.

\subsection{K Star}		

\begin{figure}[tbp] 
\epsscale{0.85}
\plotone{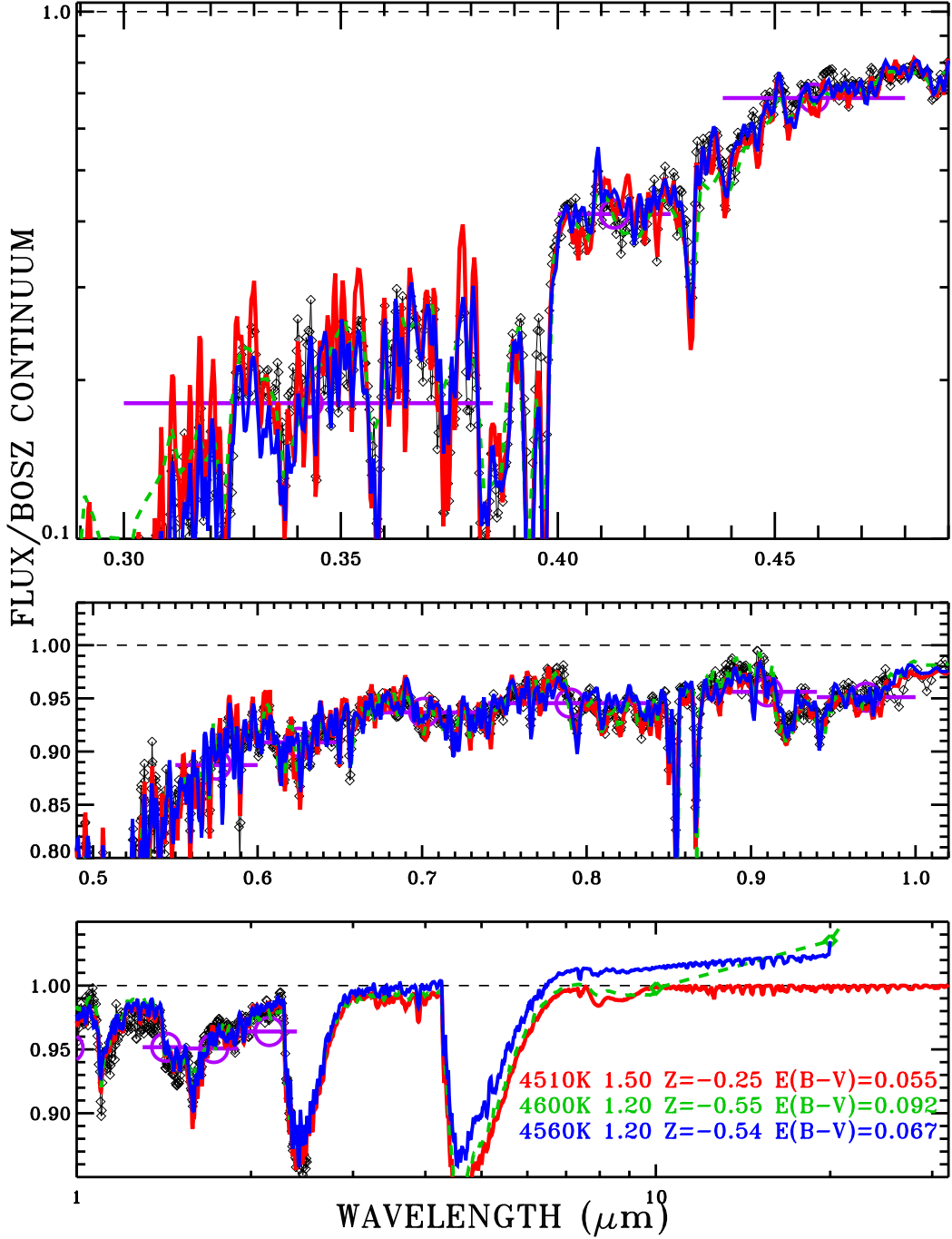}
\caption{
As in Figure~\ref{modlcont} for  $KF06T2$, where the STIS and NICMOS data are
the black diamonds connected with a thin black line, our BOSZ model is
red, CK04 is green, and Marcs is the blue line. 
Despite agreement of the data with all three
models, there is more than a 2\% disparity among the models in the IR. Notice
the strong molecular band structure in the bottom panel.
\label{kf06t2}} \end{figure}

While the proposed JWST standard stars are type G and hotter, there is
one K1.5III CALSPEC star KF06T2 (a.k.a. 2MASS J17583798+6646522). The model
fits for the BOSZ, CK04, and MARCS grids all match the measured KF06T2 fluxes
below the 2.5~\micron\ NICMOS limit but diverge by more than 2\% longward of
10~\micron, as shown Figure~\ref{kf06t2}. The precision of the IR K-star fluxes
is probably worse than for the hotter stars because of the difficulty of
modeling the strong molecular bands, especially for CO near 2.4 and 4.6~\micron.

\subsection{An Alternate Method, i.e. Fitting Balmer Lines}		

\begin{figure}			 
\centering  
\includegraphics*[width=1\textwidth,trim=0 10 0 0]{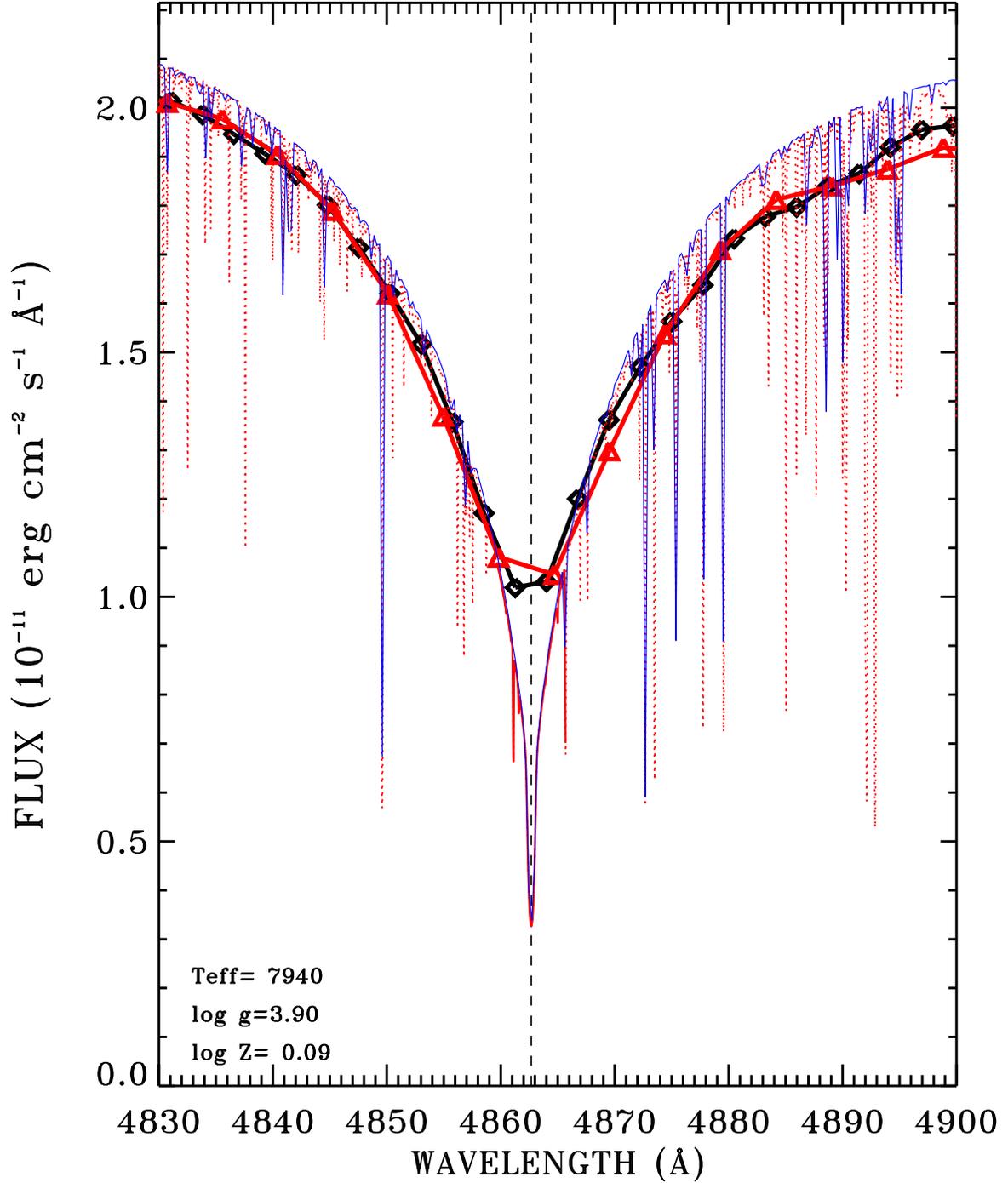}
\caption{\baselineskip=12pt
H$\beta$ region for HD14943 on a vacuum wavelength scale. Black heavy line and
diamonds are the STIS observations at a resolution R that is between 500 and 
1000, while the heavy
red line and triangles are the best fit R=500 BOSZ model with the 
Table~\ref{table:ofits} parameters that are written on the plot. The thin 
red dotted
line that is solid near line center is the BOSZ R=300,000 model, and the thin
blue line is the MARCS R=20,000 model with its best fit of
$T_\mathrm{eff}$=7930~K, $\log g$=3.90, $[M/H]$=0.07, and E(B-V)=0.012. All
models are normalized to the STIS flux at 6800--7700~\AA.
\label{balmfit}} \end{figure}

The discussion above is about fitting broad regions of the stellar continuum to
determine the $T_\mathrm{eff}$, $\log g$, and $[M/H]$. However, fitting the
Balmer line profiles provides an independent check on the temperature and
gravity, although the hydrogen line shapes are not very sensitive to
metallicity. Figure~\ref{balmfit} exemplifies the challenges of fitting high
resolution spectra for the H$\beta$ region in HD14943 (A5V). The myriad narrow
metal lines complicate automated fitting of the models to real data, as
illustrated in Figure~\ref{balmfit}. Even though the MARCS models are
monochromatic samples at R=20,000 and miss many narrow lines between sample
points \citep{plez08}, the MARCS and BOSZ show nearly perfect agreement for the overall shape
of the broad H$\beta$ line. The BOSZ and CK04 SEDs represent the mean flux in
bandpasses defined by the sample spacings. Cooler stars with narrower Balmer
lines have higher densities of confusing lines. Another caution is that the
Balmer line cores are formed high in the atmosphere, where the plane-parallel
modeling assumption may break down and render the central regions of the modeled
Balmer lines unreliable. However, the good agreement of the MARCS R=20,000
resolution with the BOSZ R=300,000 suggests that 20,000 is sufficient for a
ground-based observational program. Furthermore, the good agreement of the STIS
data with the R=500 model demonstrates that any higher resolution data should
also agree to first order with a high resolution model. For our hottest stars, a
NLTE grid of high resolution Balmer profiles would be required. For a NLTE/LTE
grid of Balmer line profiles and a discussion of the effects of stellar rotation
on the line profiles, see \citet{delgado99}. Thus, an observing program to
obtain high resolution spectra of the Balmer lines could verify our derived
$T_\mathrm{eff}$ and $\log g$, provided the stellar rotational velocity is low
and provided that the central core of the model line profiles precisely
represents the true conditions in the outer fringes of stellar atmospheres.

\section{CONCLUSIONS} 

Our new comprehensive set of model atmospheres are now publically available
through the MAST\footnote{\url{https://archive.stsci.edu/prepds/bosz}} at the
Space Telescope Science Institute and have a wavelength coverage of
0.1--32~\micron\ with an extensive variety of chemical composition, stellar
temperature, and gravity. The wavelength coverage and properly sampled SEDs with
10 spectral resolutions R in the range of 200--300,000 are essential to the flux
calibration and interpretation of stellar spectra from the James Webb Space
Telescope. No comparable sets of model spectra are publically available. Our new
BOSZ grid fits the STIS and NICMOS flux distributions well with 
$\chi^2$~=~0.1--3 for most of our stars, which have $T_\mathrm{eff}<15,000~K$. 

The CK04 LTE models for the three hottest stars have two spurious strong UV
features that are corrected in our new model grid, as illustrated for $\mu$ Col
in Figure~\ref{pltsed-bad}. One erroneous feature is excess continuum absorption
from the HeII Paschen limit at 2050.6~\AA\ (vacuum), while the extraneous
2496.8~\AA\ absorption is a single line of OIII with an overestimated line
strength.

The Lanz NLTE models are used to establish the IR SEDs of our three hottest
stars; but unfortunately, these NLTE models do not converge to their LTE
counterparts at 15,000~K, as expected. The problem may lie with an incomplete
modeling of the higher atomic hydrogen lines in the Lanz grid. However, the
complete SEDs of our cooler standards that are derived from $\chi^2$ fits of our
BOSZ models to the measured HST flux distribution have an expected 1$\sigma$
uncertainty of $\sim$1\% to their long wavelength limit of 32~\micron. The HST
fluxes at the shorter wavelengths are concatenated with their best fit R=500
BOSZ model at the longer wavelengths and are available in the CALSPEC public
database. The complete R=300,000 model SEDs also reside in
CALSPEC\footnote{http://www.stsci.edu/hst/observatory/crds/calspec.html}.
Perhaps, the most significant improvement in our predicted IR SEDs could come
from new constraints provided by Spitzer/IRAC photometry of our stars.

\acknowledgements

Primary support for this work was provided by NASA through the Space Telescope
Science Institute, which is operated by AURA, Inc., under NASA contract
NAS5-26555. Szabolcs M{\'e}sz{\'a}ros has been supported by the Premium
Postdoctoral Research Program of the Hungarian Academy of Sciences and by the
Hungarian NKFI Grants K-119517 of the Hungarian National Research, Development
and Innovation Office. A part of the computations were run on the 'atlasz'
cluster of the Eotvos Lorand University, Budapest. We acknowledge NIIF Institute
for awarding us access to resource based in Hungary at Budapest. Special thanks
to Robert Kurucz for his advice and making public his lifetime legacy of the
best LTE model atmosphere codes and and related databases. We thank Fiorella
Castelli for her guidance during the automatization of SYNTHE. This research
made use of the SIMBAD database, operated at CDS, Strasbourg, France

\newpage
\appendix
\section{Model Naming Convention}

The naming convention for the models is explained using the example
\newline amp05cp00op05t4500g05v20modrt0b500rs, where  the first letter is always "a"
meaning that the source is an ATLAS model.

\noindent mp05: abundance of metals [M/H] (mp if the [M/H] is positive, mm if 
negative.)
\newline \indent The two digits indicate the value of 10[M/H], where 
02.5 is rounded to 03,
\newline \indent 07.5  to 08, etc. only in the naming conventions.
\newline cp00: carbon abundance [C/M] (cp if the [C/M] is positive, cm 
if negative.) The 
\newline \indent digits indicate the rounded value of 10[C/M].
\newline op05: alpha element abundance [A/M] (op if [A/M] is positive, om if 
\newline \indent negative.) The digits indicate the rounded value of the 
10[A/M].
\newline t4500: The effective temperature in K.
\newline g05: The surface gravity, where the digits are the rounded value of 
the 10log g.
\newline v20: The microturbulent velocity, this is always v20, ie 2km/s.
\newline mod: parameters after this enter only the synthesis.
\newline rt0: The rotational broadening, vrot is always 0.
\newline b500: The spectral resolution R.
\newline rs: The spectra are resampled at two points per resolution element.

\begin{deluxetable}{c} 
\tablewidth{0pt}
\tablecaption{Broad Bands for Fitting G--F Star Models}
\tablehead{
\colhead{Wavelength Range (\AA)}}

\startdata
3000--3850  \\
4000--4260  \\
4380--4800  \\
4950--5500  \\
5500--6000  \\
6000--6500  \\
6620--7400  \\
7400--8400  \\
8800--9400  \\
9400--10000 \\
13000--15500\\
15500--19000\\
19000--24000\\
\enddata
\label{table:gbins} \end{deluxetable}

\begin{deluxetable}{c} 
\tablewidth{0pt}
\tablecaption{Broad Bands for Fitting A-B-O Star Models}
\tablehead{
\colhead{Wavelength Range (\AA)}}

\startdata
1280--1510 \\
1725--2020 \\
2110--2280 \\
2520--2780 \\
3000--3200 \\
3200--3400 \\
3400--3640 \\
3750--4400 \\
4400--4800 \\
4950--5500 \\
5500--6000 \\
6000--6500 \\
6620--7400 \\
7400--7900 \\
7900--8200  \\
9182--9282  \\
9290--9480  \\
9499--9599  \\
9600--10000 \\
13000--15500\\
19700--21400\\
21900--24000\\
\enddata
\label{table:obins} \end{deluxetable}

\clearpage

\bibliographystyle{apj}
\bibliography{../../pub/paper-bibliog}

\begin{deluxetable}{lccccccccccccccc}		
\rotate
\tabletypesize{\scriptsize}
\tablewidth{0pt}
\tablecaption{Parameters of the Model Fits for G Stars}
\tablehead{
\colhead{Star} &\colhead{$T_\mathrm{eff}$} &\colhead{$\log g$}
&\colhead{$[M/H]$} &\colhead{E(B-V)} &$\chi^2$ &\colhead{$T_\mathrm{eff}$} &\colhead{$\log g$}
&\colhead{$[M/H]$} &\colhead{E(B-V)} &$\chi^2$ &\colhead{$T_\mathrm{eff}$} &\colhead{$\log g$}
&\colhead{$[M/H]$} &\colhead{E(B-V)} &$\chi^2$ \\
&\multicolumn{1}{r}{BOSZ} &&&&&\multicolumn{0}{r}{CK04} &&&&&\multicolumn{0}{r}{MARCS}
}
\startdata
                      C26202  &6300  &4.65 &-0.39 &0.069  &0.51    &6320  &4.60 &-0.45  &0.073  &0.48    &6320  &4.85 &-0.49  &0.076  &0.52\\
                     HD37962  &5750  &3.75 &-0.20 &0.012  &0.28    &5690  &3.35 &-0.41  &0.002  &0.18    &5790  &4.35 &-0.31  &0.024  &0.35\\
                     HD38949  &5990  &4.30 &-0.11 &0.001  &0.08    &5980  &4.00 &-0.25  &0.001  &0.07    &5990  &4.60 &-0.26  &0.005  &0.18\\
                    HD106252  &5830  &3.85 &-0.11 &0.000  &0.17    &5820  &3.50 &-0.28  &0.000  &0.16    &5810  &4.20 &-0.27  &0.000  &0.21\\
      P041C\tablenotemark{a}  &5960  &4.15 & 0.11 &0.022  &0.17    &6040  &4.00 & 0.05  &0.040  &0.30    &6030  &4.70 & 0.02  &0.040  &0.22\\
                       P177D  &5850  &3.65 &-0.02 &0.045  &0.21    &5880  &3.45 &-0.14  &0.053  &0.31    &5910  &4.20 &-0.10  &0.061  &0.26\\
                 SF1615+001A  &5820  &4.45 &-0.61 &0.105  &0.43    &5880  &4.20 &-0.69  &0.118  &0.60    &5860  &4.85 &-0.77  &0.116  &0.74\\
                       SNAP2  &5750  &4.40 &-0.19 &0.033  &0.11    &5810  &4.10 &-0.28  &0.047  &0.12    &5800  &4.75 &-0.32  &0.048  &0.26\\
                       P330E  &5840  &4.40 &-0.16 &0.036  &0.32    &5900  &4.10 &-0.25  &0.049  &0.49    &5900  &4.75 &-0.29  &0.052  &0.52\\
                    HD159222  &5800  &3.55 & 0.08 &0.000  &0.16    &5790  &3.30 &-0.09  &0.001  &0.12    &5780  &4.10 &-0.07  &0.001  &0.19\\
                    HD205905  &5850  &3.75 & 0.03 &0.003  &0.18    &5830  &3.45 &-0.14  &0.001  &0.14    &5870  &4.30 &-0.08  &0.011  &0.18\\
   HD209458\tablenotemark{b}  &6090  &4.15 & 0.01 &0.002  &0.13    &6160  &4.05 &-0.04  &0.017  &0.18    &6150  &4.55 &-0.07  &0.017  &0.20\\
\enddata
\tablenotetext{a}{P041C has an M companion 0.57arcsec away 
	\citep{gilliland11}}
\tablenotetext{b}{Transiting planet. See B10.}
\label{table:gfits} \end{deluxetable}

\begin{deluxetable}{lcccccccccc}		
\tabletypesize{\scriptsize}
\tablewidth{0pt}
\tablecaption{Parameters of the Model Fits for OBA Stars}
\tablehead{
\colhead{Star} &\colhead{$T_\mathrm{eff}$} &\colhead{$\log g$}
&\colhead{$[M/H]$} &\colhead{E(B-V)} &$\chi^2$ &\colhead{$T_\mathrm{eff}$} &\colhead{$\log g$}
&\colhead{$[M/H]$} &\colhead{E(B-V)} &$\chi^2$  \\
&\multicolumn{1}{r}{BOSZ} &&&&&\multicolumn{0}{r}{CK04}
}
\startdata
10 Lac             &30950 & 3.90 &0.00 & 0.075 &3.27 &30910 &3.95 & 0.10 & 0.077 &2.92 \\
$\lambda$ Lep      &27080 & 3.80 &-0.28 & 0.005 &3.52 &27100 &3.90 &-0.22 & 0.006 &3.40 \\
$\mu$ Col          &30980 & 4.25 & 0.20 & 0.009 &3.52 &30950 &4.35 & 0.32 & 0.011 &3.34 \\
$\xi^{2}$  Cet     &10370 & 3.95 &-0.62 & 0.000 &2.85 &10370 &3.95 &-0.52 & 0.000 &1.97  \\
HD014943           & 7940 & 3.90 & 0.09 & 0.011 &0.76 & 7930 &3.90 & 0.07 & 0.012 &1.04  \\
HD37725            & 8350 & 4.25 &-0.10 & 0.041 &1.17 & 8380 &4.30 &-0.08 & 0.045 &1.43  \\
HD116405           &10790 & 4.00 &-0.35 & 0.000 &0.61 &10790 &4.05 &-0.37 & 0.000 &0.42  \\
BD+60$^{\circ}$1753& 9370 & 3.90 &-0.09 & 0.013 &0.84 & 9410 &3.90 &-0.06 & 0.017 &1.01  \\
HD158485           & 8580 & 4.15 &-0.39 & 0.046 &1.47 & 8640 &4.20 &-0.35 & 0.052 &1.72  \\
1732526            & 8630 & 4.10 &-0.32 & 0.036 &2.92 & 8670 &4.15 &-0.25 & 0.039 &2.99  \\
1743045            & 7470 & 3.65 &-0.29 & 0.026 &1.38 & 7460 &3.65 &-0.31 & 0.026 &1.59 \\
HD163466           & 7960 & 3.75 &-0.21 & 0.031 &2.21 & 7950 &3.75 &-0.24 & 0.031 &2.53  \\
1757132            & 7640 & 3.75 & 0.19 & 0.036 &0.83 & 7660 &3.80 & 0.18 & 0.041 &1.23 \\
1802271            & 9040 & 4.00 &-0.48 & 0.017 &0.94 & 9070 &4.00 &-0.47 & 0.020 &0.98 \\
1805292            & 8570 & 4.00 &-0.07 & 0.034 &0.80 & 8540 &4.00 &-0.11 & 0.032 &0.90 \\
1808347            & 7910 & 3.85 &-0.61 & 0.024 &2.46 & 7890 &3.85 &-0.62 & 0.022 &2.80 \\
1812095            & 7810 & 3.65 & 0.22 & 0.008 &0.88 & 7830 &3.70 & 0.22 & 0.013 &0.95 \\
HD180609           & 8560 & 3.95 &-0.44 & 0.037 &0.59 & 8600 &4.00 &-0.45 & 0.042 &0.75 \\
HD165459\tablenotemark{a} & 8570 & 4.20 & 0.10 &0.023 &0.53 & 8540 &4.20 & 0.07 & 0.021 &0.64 \\
\enddata
\tablenotetext{a}{There is a dust ring that affects the SED longward of 
8~\micron.
See \citet{bohlin11}.}
\label{table:ofits} \end{deluxetable}

\begin{deluxetable}{lccccc}		
\tabletypesize{\scriptsize}
\tablewidth{0pt}
\tablecaption{Parameters of the Lanz \& Hubeny Model Fits for O Stars}
\tablehead{
\colhead{Star} &\colhead{$T_\mathrm{eff}$} &\colhead{$\log g$}
&\colhead{$[M/H]$} &\colhead{E(B-V)} &$\chi^2$  \\
}
\startdata
10 Lac        &32190 & 3.65 & 0.05 &0.071 &3.20 \\
$\lambda$ Lep &27170 & 3.25 &-0.03 &0.003 &1.94 \\
$\mu$ Col     &31640 & 3.65 & 0.13 &0.001 &1.76 \\
\enddata
\label{table:lanz} \end{deluxetable}

\end{document}